\documentclass[twocolumn,floatfix,prb,aps,showpacs,superscriptaddress]{revtex4-1}
\usepackage{graphicx,amsmath,amssymb,color}
\usepackage{nicefrac}
\usepackage{dsfont}
\usepackage[titletoc,title]{appendix}
\usepackage{hyperref}

\newcommand{\be}{\begin{equation}}
\newcommand{\ee}{\end{equation}}

\newcommand{\ba}{\begin{eqnarray}}
\newcommand{\ea}{\end{eqnarray}}

\def\bea{\begin{eqnarray}}
\def\eea{\end{eqnarray}}
\def\ba{\begin{equation}\begin{array}{c}}
\def\ea{\end{array}\end{equation}}
\def\be{\ba\displaystyle}
\def\ee{\ea}

\begin{document}

\title{
The Enigma
of the $\nu=2+3/8$ Fractional Quantum Hall Effect}
\author{Jimmy A. Hutasoit}
\affiliation{Instituut-Lorentz, Universiteit Leiden, P.O. Box 9506, 2300 RA Leiden, The Netherlands}

\author{Ajit C. Balram}
\affiliation{Department of Physics, 104 Davey Lab, Pennsylvania State University, University Park, PA 16802, USA}
\affiliation{Niels Bohr International Academy and the Center for Quantum Devices, 
Niels Bohr Institute, University of Copenhagen, 2100 Copenhagen, Denmark}

\author{Sutirtha Mukherjee}
\affiliation{Department of Theoretical Physics, Indian Association for the Cultivation of Science, Jadavpur, Kolkata 700032, India}

\author{Ying-Hai Wu}
\affiliation{Max-Planck-Institut f{\"u}r Quantenoptik, Hans-Kopfermann-Stra{\ss}e 1, 85748 Garching, Germany}

\author{Sudhansu S. Mandal}
\affiliation{Department of Physics, Indian Institute of Technology Kharagpur, Kharagpur 721 302, India}  

\author{A. W\'ojs}
\affiliation{Department of Theoretical Physics, Wroc\l{}aw University of Science and Technology, 50-370 Wroc\l{}aw, Poland}  

\author{Vadim Cheianov}
\affiliation{Instituut-Lorentz, Universiteit Leiden, P.O. Box 9506, 2300 RA Leiden, The Netherlands}

\author{J. K. Jain}
\affiliation{Department of Physics, 104 Davey Lab, Pennsylvania State University, University Park, PA 16802, USA}
\affiliation{Department of Physics, Indian Institute of Science, Bengaluru, 560012, India}

\begin{abstract} 
The fractional quantum Hall effect at $\nu=2+3/8$, which has been definitively observed, is one of the last fractions for which no viable explanation has so far been demonstrated. Our detailed study suggests that it belongs to a new class of of exotic states described by the Bonderson-Slingerland wave function. Its excitations are non-Abelian anyons similar to those of the well studied Pfaffian state at 5/2, but its wave function has a more complex structure. Using the effective edge theory, we make predictions for various measurable quantities that should enable a confirmation of the underlying topological order of this state.
\end{abstract}
\maketitle 

\section{Introduction}
The 5/2 fractional quantum Hall effect (FQHE), the first even denominator FQH state (FQHS) to be observed \cite{Willett87}, has produced a host of remarkable concepts. Its most plausible explanation is in terms of the Moore-Read (MR) Pfaffian (Pf) wave function \cite{Moore91}, which represents a topological $p$-wave paired state of fully spin polarized composite fermions (CFs) \cite{Jain89} supporting Majorana modes with non-Abelian braid statistics \cite{Read00}.  Composite fermions with weak repulsive interaction form a Fermi sea at a half filled Landau level (LL) \cite{Halperin93}, as is the case at $\nu=1/2$ in the lowest LL (LLL), but presumably, the interaction between them is weakly attractive at $\nu=5/2$, leading to the formation of CF-pairs, which opens a gap and produces an incompressible state \cite{Scarola00b}.  The particle-hole conjugate of the MR Pfaffian state, called the anti-Pfaffian (APf) \cite{Levin07,Lee07}, is a topologically distinct candidate for the 5/2 FQHE, and a breaking of the particle-hole symmetry due to LL mixing will determine which of the two is favored in experiments \cite{Bishara09,Wang09,Wojs10,Rezayi11,Peterson14,Zaletel15,Pakrouski15,Tyler15}. 

This article is concerned with the physical origin of another even denominator fraction, namely $\nu=2+3/8$ for which definitive experimental evidence exists \cite{Xia04,Pan08,Choi08,Kumar10,Zhang12}; notably, Kumar {\em et al.} \cite{Kumar10} have reported activated behavior at this filling fraction establishing incompressibility.  The underlying physical mechanism of the $2+3/8$ FQHE has not yet been understood \cite{Toke08}. We  consider three topologically distinct candidates for this state, namely the Pfaffian and the anti-Pfaffian states of composite fermions, and a Bonderson-Slingerland (BS) state, all defined below, and find that it is best described by the BS state. All of these states also support excitations with non-Abelian braid statistics \cite{Bonderson08}. We then consider measurements that can confirm the BS topological order of the $2+3/8$ state. 

Other states have also been proposed as candidates for the 3/8 FQHE. A maximally chiral Abelian $3/8$ state was obtained in the effective-edge-theory based classification of Fr{\"o}hlich {\em et al.} \cite{Frohlich97} but without a prescription for a trial wave function. Jolicoeur proposed a series of non-Abelian wave functions at $\nu=k/(3k-2)$ \cite{Jolicoeur07}, but the Jolicoeur 3/8 wave function ($k=6$) is not readily amenable to evaluation.

\section{Trial wave functions}

We discuss the wave functions in Haldane's spherical geometry \cite{Haldane83} where $N$ electrons reside on the surface of a sphere and a magnetic monopole of strength $2Q\phi_{0}$ (where $\phi_0=hc/e$ is a flux quantum) produces a radial magnetic field. For finite systems, the incompressible state at a filling factor $\nu$ occurs at $2Q=\nu^{-1}N-\mathcal{S}$, where $\mathcal{S}$ is called the shift. Topologically distinct candidate states often occur at different shifts. Unless otherwise stated we shall assume the electrons to be fully spin polarized, which is consistent with the current experimental results. We shall also discuss below all physics within the LLL subspace, even though we are interested in the second LL physics. This is possible because the problem of electrons in the second LL interacting via the Coulomb interaction is mathematically equivalent to the problem of electrons in the LLL interacting with an effective interaction that has the same Haldane pseudopotentials \cite{Haldane83} as the Coulomb interaction in the second LL. We will neglect the effects of LL mixing and finite width, which alter the form of the interaction and produce corrections to various observable quantities, but are not expected to change the form of correlations and thus, will not destroy the microscopic mechanism for incompressibility we propose here. Nevertheless, such studies will be needed for a more reliable quantitative comparison with experiments. 

The filling factor of composite fermions, $\nu^*$, is related to the electron filling factor by the relation $\nu=\nu^*/(2p\nu^*\pm 1)$ \cite{Jain89,Jain07}. If composite fermions carrying two vortices ($p=1$) are formed at $\nu=3/8$ in the second LL (taking the lowest filled LL as inert), their filling factor is $\nu^*=3/2$. The most immediate possibility one can imagine is one in which composite fermions in the second $\Lambda$ level (Landau-like level of composite fermions) capture two more vortices and form a MR-like paired Pfaffian or anti-Pfaffian state. The Pf and APf trial wave functions for $\nu=3/8$ can be constructed by composite-fermionizing the 3/2 Pfaffian or the anti-Pfaffian wave function as follows \cite{Mukherjee12}:
\begin{eqnarray}
\Psi_{3/8}^{\text{APf/Pf}}&=&\mathcal{P}_{\text{LLL}} \prod_{j<k} (u_{j}v_{k}-u_{k}v_{j})^{2} \Phi_{3/2}^{\text{APf/Pf}} 
\label{eq:trial}
\end{eqnarray}
where $u=\cos({\theta/2})\exp({i\phi/2})$ and $v=\sin({\theta/2})\exp({-i\phi/2})$ are the spinor coordinates on the sphere, $\mathcal{P}_{\text{LLL}}$ implements LLL projection and $\Phi_{3/2}^{\text{APf/Pf}}$ refers to the APf or Pf state at filling factor 3/2, in which the lowest LL is fully filled and electrons in the second LL form a paired APf or Pf state. The Pf state in the LLL is obtained by diagonalizing a three-body interaction Hamiltonian given by \cite{Greiter91,Greiter92a}: $V_{3}^{\text{int}}=\sum_{i<j<k} \mathcal{P}_{ijk}^{(3)}(3Q-3)$ where $\mathcal{P}_{ijk}^{(3)}(L)$ is the projection operator which projects the state of three particles $i$, $j$ and $k$ into the subspace of total orbital angular momentum $L$. The APf state is obtained by taking the particle-hole conjugate of the Pf state. $\Phi_{3/2}^{\text{APf/Pf}}$ is obtained by elevating these states to the second LL and fully populating the LLL. 

The MR Pf and APf states at 1/2 occur at $2Q=2N-3$ and $2Q=2N+1$, respectively. The corresponding Pf and APf trial wave functions $\Psi_{3/8}^{\text{APf/Pf}}$ at $\nu=3/8$ occur at $2Q=(8N-13)/3$ and $2Q=(8N-9)/3$ (shifts $\mathcal{S}=13/3$ and $\mathcal{S}=3$), respectively. The largest systems accessible to exact diagonalization are $N=14$ for the shift $\mathcal{S}=13/3$ and $N=12$ for $\mathcal{S}=3$. In spite of the large Hilbert space dimensions of these systems (Table  \ref{tab:overlaps}), it is possible to obtain the overlaps of these states with the exact Coulomb ground state in the second LL at the corresponding flux. These overlaps, shown in Table \ref{tab:overlaps}, are low ($<10\%$). We have also considered a slightly modified trial wave function in which $\Phi_{3/2}^{\text{APf/Pf}}$ on the right hand side of Eq.~\ref{eq:trial} is replaced by the exact Coulomb state at 3/2 at the appropriate flux (for fully spin polarized particles). The overlaps of the resulting wave function with the corresponding exact Coulomb ground state are also given in Table \ref{tab:overlaps} (denoted overlap$'$), and are also very low. We believe that the rather poor overlaps rule out the Pf or APf pairing of composite fermions in the second $\Lambda$ level as the origin of the 3/8 FQHE in the second LL. 

These results are in stark contrast with the 3/8 FQHE in the LLL. Mukherjee {\em et al.} \cite{Mukherjee12} showed the APf wave function to be plausible for the fully spin polarized FQHE at this fraction. We note, however, that while experimental signatures for the 3/8 state have been seen \cite{Pan03,Bellani10}, its existence has not yet been confirmed \cite{Pan15,Samkharadze15b}.

\begin{table}
\begin{center}
\begin{tabular}{|c|c|c|c|c|c|c|}
\hline
$N$ 		& 2Q		& $\text{dim}_{L=0}$	& $\text{dim}_{L_{z}=0}$& trial state 	& overlap	& overlap$'$	\\ \hline
12		& 29		&1,330& 1,437,269& $\Psi_{3/8}^{\rm APf}$ 	&	0.091(2)	&	0.083(1)	\\ \hline
14		& 33		&  11,463 & 19,159,798	& $\Psi_{3/8}^{\rm Pf}$		&	0.074(3)	&	0.104(1)	\\ \hline
12		& 31		& 2,634	&3,451,798 	&$\Psi_{3/8}^{\rm BS}$		&	0.85442(51)	&	- 		\\ \hline
\end{tabular}
\end{center}
\caption {Overlaps of the trial wave functions described in Eq. \ref{eq:trial} and \ref{eq_BS_3_8} with the exact  Coulomb ground state at $3/8$ in the second LL. The last column shows the overlaps with the trial state obtained by composite-fermionizing the exact Coulomb ground states at $3/2$ at the appropriate effective flux values. The total Hilbert space dimension ($\text{dim}_{L_{z}=0}$) and the zero total orbital angular momentum subspace dimension ($\text{dim}_{L=0}$) are also shown. The number in the parenthesis is the statistical error in the Monte Carlo evaluation of the overlap.}  
\label{tab:overlaps} 
\end{table}

Having ruled out the Pf / APf pairing of composite fermions, we consider another candidate state proposed by BS, who have constructed a class of wave functions that generalize the MR Pfaffian wave function \cite{Bonderson08}. Just as the MR wave function is obtained by multiplying the bosonic Laughlin 1/2 state by the Pfaffian factor, the BS wave functions are obtained by multiplying the bosonic Jain states at $\nu=n/[(2p+1)n\pm 1]$ \cite{Regnault03,Chang05b} by the Pfaffian factor:
\begin{eqnarray}
\Psi^{\rm BS}_{n \over (2p+1)n\pm 1}&=&{\rm Pf}\left[{1\over u_{j}v_{k}-u_{k}v_{j}} \right] \prod_{j<k}(u_{j}v_{k}-u_{k}v_{j})^{2p} \nonumber \\
&&\times~{\cal P}_{\rm LLL} \prod_{j<k}(u_{j}v_{k}-u_{k}v_{j})\Phi_{\pm n}
\end{eqnarray}
where $\Phi_n$ is the wave function of $n$ filled LLs of electrons, and $\Phi_{-n}\equiv [\Phi_n]^*$ represents $n$ filled LLs in negative fields. For $p=0$, the only acceptable state is the MR Pfaffian state corresponding to $n=1$ and $+$ sign in the above equation; other states of the form ${\cal P}_{\rm LLL} \prod_{j<k}(u_{j}v_{k}-u_{k}v_{j})\Phi_{\pm n}$ have non-zero probability of two particles being at the same point, producing an ill-defined wave function when multiplied by the Pfaffian factor. For $2p=2$ on the other hand, all BS wave functions are well defined. In particular, the BS wave function for the 3/8 state is given by (in a form that is most suitable for our calculations):
\begin{equation}
\Psi_{{3/8}}^{\text{BS}}= \text{Pf}~\Phi_{1}\mathcal{P}_{\text{LLL}} \Phi_{1}^2[\Phi_{3}]^{*} = \text{Pf}~\Phi_{1} \Psi_{3/5}
\label{eq_BS_3_8}
\end{equation}
where $\Phi_{1}=\prod_{j<k}(u_{j}v_{k}-u_{k}v_{j})$ is the wave function of filled LLL, $\text{Pf}={\rm Pf}\left[{1\over u_{j}v_{k}-u_{k}v_{j}} \right]$, and $\Psi_{3/ 5}$ is the Jain $3/5$ state \cite{Jain89}, with the LLL projection conveniently evaluated following Refs. \cite{Jain97,Jain97b,Davenport12}. $\Psi_{{3}/{8}}^{\text{BS}}$ occurs at a shift of $\mathcal{S}=1$, i.e. at $2Q=8N/3-1$. 

\begin{figure}
\begin{center} 
\includegraphics[width=3.2in, height=1.9in]{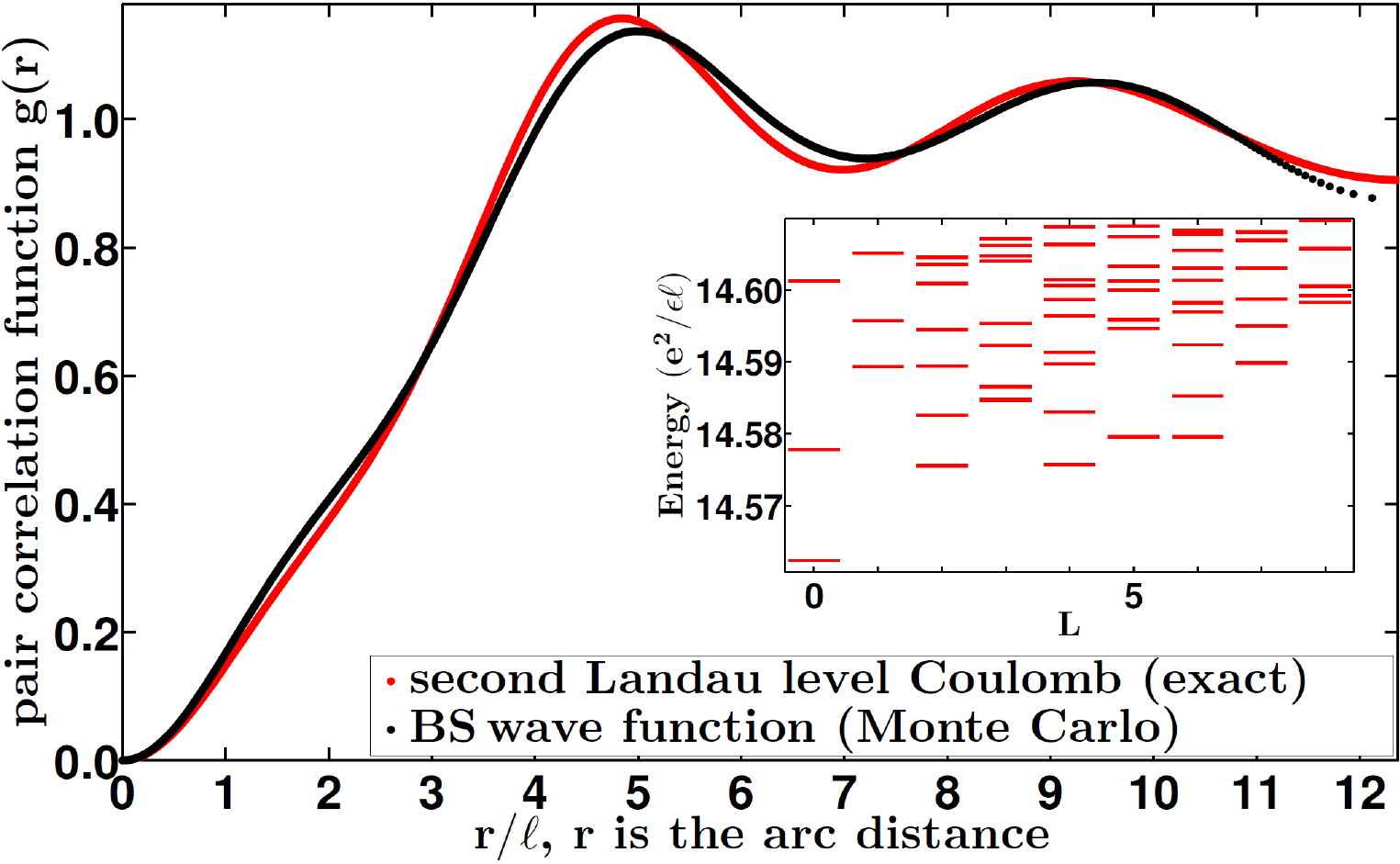}
\end{center}  
\caption {The pair correlation function $g(r)$ for the exact second Landau level Coulomb state (red), and the Bonderson-Slingerland trial state of Eq. \ref{eq_BS_3_8} (black) at filling factor $\nu=2+3/8$ for $N=12$ electrons. Inset: Spectrum for the fully polarized $\nu=3/8$ state in the second LL obtained by exact diagonalization at $N=12$ and $2Q=31$ for the exact second Landau level Coulomb interaction.
}
\label{fig:5} 
\end{figure}

The exact Coulomb ground state for $N=12$ particles at shift $\mathcal{S}=1$ is incompressible, see the inset of Fig. \ref{fig:5}. Its overlap with the model BS wave function is 0.85442(51). It is much larger than the overlaps for the $\Psi_{3/8}^{\rm Pf}$ or the $\Psi_{3/8}^{\rm APf}$, and roughly of the same order as the overlaps of the MR Pf wave function with the exact 5/2 state \cite{Morf98,Scarola02} and those of the 1/3 Laughlin wave function with the exact 7/3 state \cite{Balram13b}.  (For reference, for $N=10$ and 14,  the overlap of the Pfaffian wave function with the exact 5/2 Coulomb ground state are 0.84 and 0.69, whereas the overlaps of the Laughlin 1/3 state with the exact 7/3 state are 0.54 and 0.58  \cite{Morf98,Scarola02,Wojs09}.) Exact diagonalization for the next system size ($N=18,~2Q=47$) is beyond the reach of our current computational capability. From the spectrum of $N=12$, we estimate the gap to be of the order of $0.01$ $e^2/\epsilon\ell$ which is close to the experimental estimated value $0.015$ $e^2/\epsilon\ell$ \cite{Kumar10}. 
(The theoretical gap for the LLL 3/8 is $\sim 0.002$ $e^2/\epsilon\ell$ \cite{Mukherjee12}.) Fig. \ref{fig:5} compares the pair correlation functions $g(r)$ of the exact Coulomb ground state with that of the BS state. These match very well and show decaying oscillations at long distances, typical of an incompressible state. We also see a ``shoulder'' like feature (also seen in the LLL 3/8 \cite{Wojs05}, MR \cite{Read96,Park96}, Read-Rezayi (RR) \cite{Read99,Rezayi09} and other BS states) at short distances which is considered to be a characteristic fingerprint of non-Abelian states \cite{Bonderson12} that involve pairing (MR) or clustering (RR at $\nu=k/(k+2)$ is a $k$-cluster). These results taken together give us a fair degree of confidence in the validity of the BS wave function for the $2+3/8$ FQHE.

We have also considered the possibility of a partially polarized $2+3/8$ FQHE state. All three states considered above have partially spin polarized cousins. The partially polarized APf/Pf states are given by ${\cal P}_{\rm LLL}\Phi_1^2[\Phi_{1,\uparrow}\Psi^{\rm APf/Pf}_{1/2,\downarrow}]$, and the partially polarized BS state is given by 
$\text{Pf}~ \mathcal{P}_{\text{LLL}} \Phi_{1}^{3}[\Phi_{2,\uparrow}\Phi_{1,\downarrow}]^{*}$.  Exact diagonalization studies up to $N=10$ do not support the realization of any of these states for spinful electrons at zero Zeeman energy (the ground states at the appropriate $N$ and $2Q$ values do not have the predicted quantum numbers).  
This should be contrasted with the situation at $\nu=3/8$ in the LLL, where a partially spin polarized 3/8 state is predicted to occur at sufficiently low Zeeman energies\cite{Scarola02b,Mukherjee14c,Balram15}, and to provide an almost exact representation for the APf pairing of composite fermions \cite{Mukherjee14c}.

\section{Energy Spectrum on Torus}

In addition to wave function overlap, we can also study the energy spectrum of this state on torus. We consider a rectangular torus spanned by vectors $\mathbf{L}_{1}=L_{1}{\widehat e}_{x},\mathbf{L}_{2}=L_{2}{\widehat e}_{y}$ and we choose the Landau gauge ${\mathbf A}=(0,Bx,0)$. The aspect ratio of the torus is defined as $R=L_{1}/L_{2}$. For a torus enclosing a total magnetic flux of $N_{\phi}$, the single-particle wave functions in the second Landau level are
\begin{widetext}
\begin{eqnarray}
\psi^{N_{\phi}}_{m}(x,y) = \frac{1}{(2L_{2}\ell_{B}\sqrt{{\pi}})^{1/2}} && \sum^{\mathbb{Z}}_{k} H_{1} \left[ \frac{x}{\ell_B} - \frac{2{\pi}\ell_{B}}{L_2} \left( m+kN_{\phi} \right) \right] \nonumber \\
&\times& \exp \left\{ - \frac{1}{2} \left[ \frac{x}{\ell_B} - \frac{2{\pi}\ell_{B}}{L_2} \left( m+kN_{\phi} \right) + i \frac{2{\pi}y}{L_2} \left( m+kN_{\phi} \right) \right]^2 \right\},
\end{eqnarray}
\end{widetext}
where $\ell_{B}=\sqrt{L_{1}L_{2}/(2{\pi}N_{\phi})}$ is the magnetic length and $H_n$ denotes the Hermite polynomial of degree $n$. The creation (annihilation) operator for the single-particle state with quantum number $m$ is denoted as $C^\dagger_{m}$ ($C_{m}$). The electrons interact via Coulomb potential $V({\mathbf r}_1-{\mathbf r}_2)=e^2/(\varepsilon|{\mathbf r}_1-{\mathbf r}_2|)$. The second quantized form of the many-body Hamiltonian is
\begin{eqnarray}
\frac{1}{2} \sum_{\{m_{i}\}} F_{m_{1}m_{2}m_{4}m_{3}} C^\dagger_{m_{1}} C^\dagger_{m_{2}} C_{m_{4}} C_{m_{3}}.
\label{ManyHamilton}
\end{eqnarray}
By defining the reciprocal lattice vectors ${\mathbf G}_{1}=2\pi{\widehat e}_{x}/L_{1},{\mathbf G}_{2}=2\pi{\widehat e}_{y}/L_{2}$, we transform the interaction potential to momentum space as 
\begin{eqnarray}
V({\mathbf r}_1 - {\mathbf r}_2) &=& \frac{1}{L_{1}L_{2}} \sum_{\mathbf{q}} V(\mathbf q) e^{i{\mathbf q}\cdot({\mathbf r}_1 - {\mathbf r}_2)},
\end{eqnarray}
where ${\mathbf q}=q_1{\mathbf G}_1+q_2{\mathbf G}_2$. This helps us to find that the coefficients $F_{m_1m_2m_4m_3}$ are
\begin{widetext}
\begin{eqnarray}
&& \int d^2 {\mathbf r}_1 d^2 {\mathbf r}_2 \; \left[ \psi^{N_{\phi}}_{m_1}({\mathbf r}_1) \right]^* \left[ \psi^{N_{\phi}}_{m_2}({\mathbf r}_2) \right]^* V({\mathbf r}_1-{\mathbf r}_2) \psi^{N_{\phi}}_{m_4}({\mathbf r}_2) \psi^{N_{\phi}}_{m_3}({\mathbf r}_1) \nonumber \\
= && \frac{1}{N_{\phi}} \sum^{N_{\phi}}_{m_1} \sum^{N_{\phi}}_{m_2} \sum_{q_1,q_2} V({\mathbf q}) \exp \left\{ -\frac{1}{2} {\mathbf q}^2 \ell^{2}_{B} \right\} \left( 1 - \frac{{\mathbf q}^2}{2} \right)^2 \exp \left\{ i2{\pi}q_1 \left[ \frac{m_1-m_2-q_2}{N_{\phi}} \right] \right\} {\widetilde\delta}^{N_{\phi}}_{m_1+m_2,m_3+m_4},
\end{eqnarray} 
\end{widetext}
where ${\widetilde\delta}^{N_{\phi}}_{i,j}$ is a generalized Kronecker delta defined as ${\widetilde\delta}^{N_{\phi}}_{i,j}=1 \;\; {\rm iff} \;\; i \; {\rm mod} \; N_{\phi} = j \; {\rm mod} \; N_{\phi}$. The many-body eigenstates are labeled by the momentum quantum number $Y \equiv ( \sum^{{N}_{\phi}}_{i=1} m_{i} ) \; {\rm mod} \; N_{\phi}$.

\begin{figure}
\includegraphics[width=0.45\textwidth]{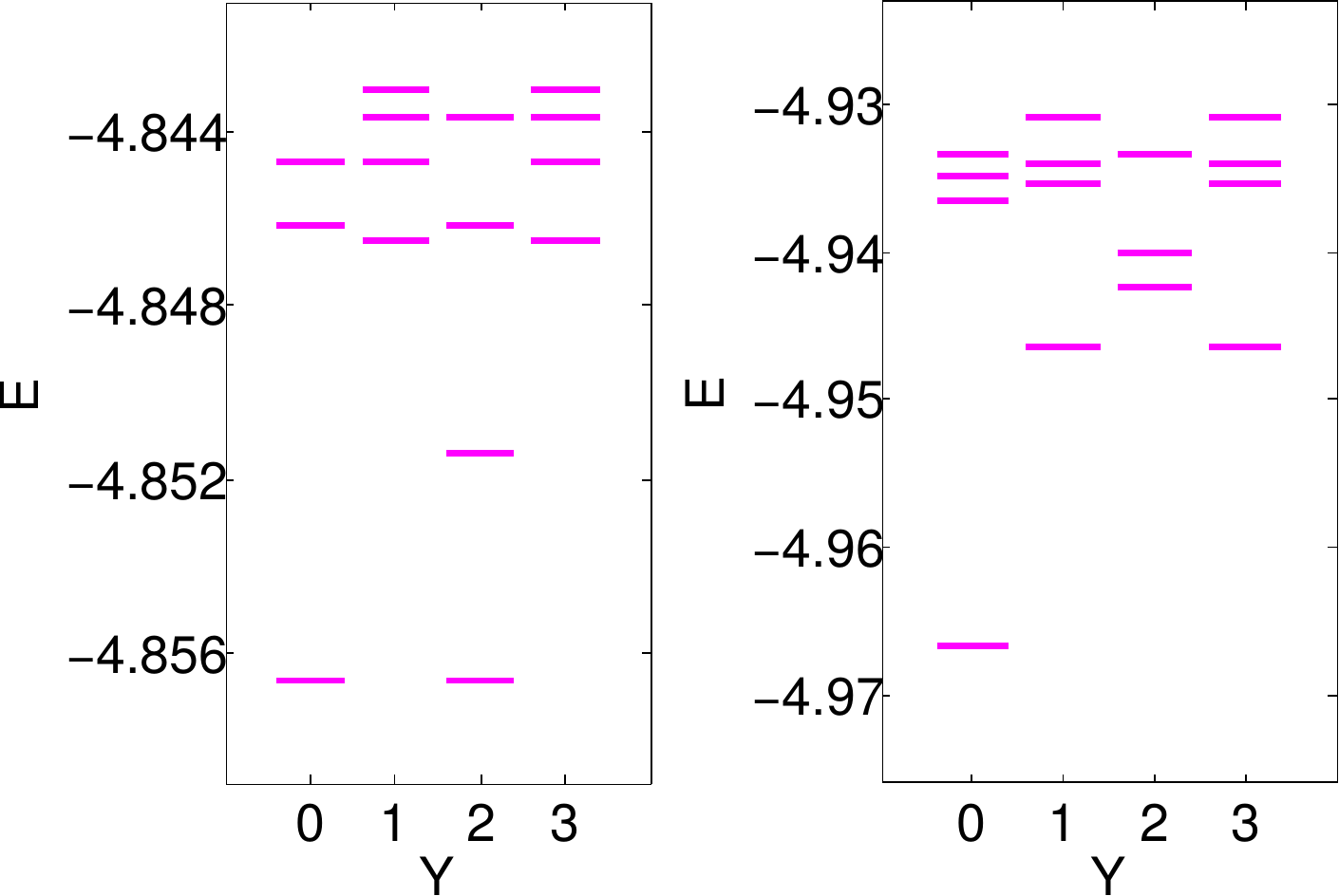}
\caption{Energy spectra of the $N=12,N_{\phi}=32$ system on torus with aspect ratio (a) $R=1.0$ and (b) $R=1.5$.}
\label{FigureS1}
\end{figure}

One characteristic feature of topologically ordered states is the existence of multiple degenerate ground states on torus. The ground state degeneracy of the BS state at $\nu=3/8$ is expected to be $24$ based on the following argument: there is a center of mass degeneracy $q$ for a system at $\nu=p/q$ \cite{Haldane85b} (which is exact for any translationally invariant Hamiltonian) and the Pfaffian factor gives rise to another three fold degeneracy in each center of mass sector (which is not exact in finite size systems but converges in the thermodynamic limit). We show in Fig. \ref{FigureS1} the energy spectra of the $N=12,N_{\phi}=32$ system on torus in one center of mass sector (the other sectors are simply duplicates of this one). One can see evidence for that there is a three fold quasi degeneracy for $R=1.0$ but such degeneracy is not robust as it disappears at $R=1.5$.

\section{Entanglement Spectrum on Sphere}

\begin{figure}
\includegraphics[width=0.45\textwidth]{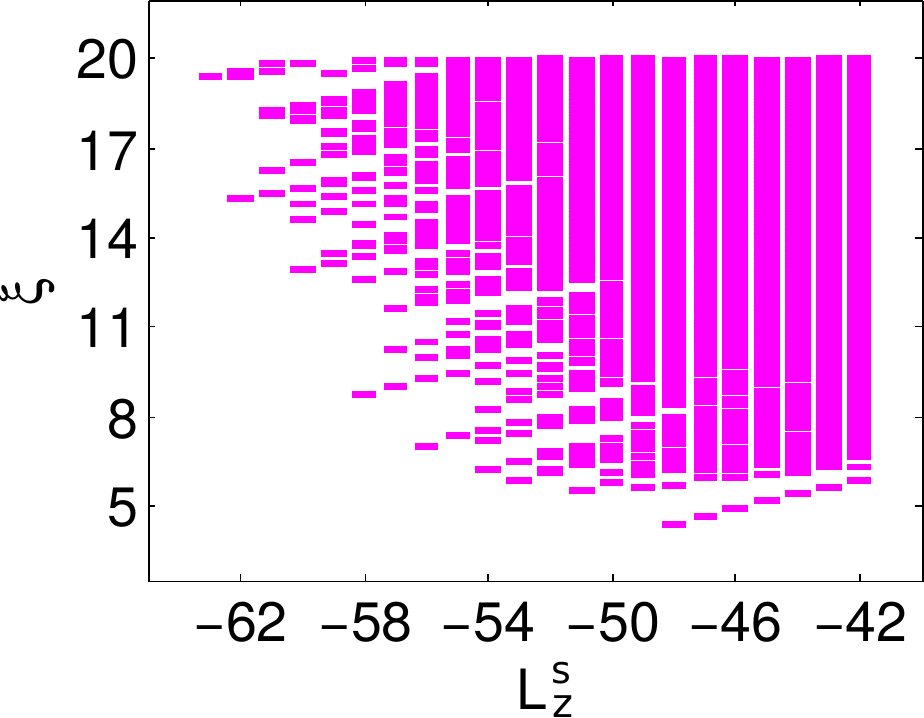}
\caption{Entanglement spectrum of the $N=12,2Q=31$ system on sphere. $S$ ($R$) is the southern (northern) hemisphere. The number of electrons in the southern hemisphere is $N_{A}=6$. The entanglement levels are labeled by the $z$-component angular momentum of the southern hemisphere $L^{s}_{z}$.}
\label{FigureS2}
\end{figure}

The entanglement spectrum has been useful in characterizing some FQH states \cite{Li08}. To compute it from a ground state $|\Psi\rangle$, we divide the system into two subsystems $S$ and $R$ so the state is decomposed as $|\Psi\rangle = \sum_{ij} F_{ij} |\Psi^{S}_{i}\rangle \otimes |\Psi^{R}_{j}\rangle = \sum_{\mu} e^{-\xi_{\mu}/2} |\Psi^{S}_{\mu}\rangle \otimes |\Psi^{R}_{\mu}\rangle$, where $|\Psi^{S}_{i}\rangle$ and $|\Psi^{R}_{j}\rangle$ are the basis states of $S$ and $R$ respectively. The Schmidt decomposition of matrix $F_{ij}$ gives entanglement eigenvalues $\xi_{\mu}$. We use the ground state on sphere and cut the sphere along its equator to create a virtual edge so one may extract some information about the physical edge states \cite{Dubail12,Sterdyniak12,Rodriguez12}. Fig. \ref{FigureS2} shows the entanglement spectrum of the $N=12, 2Q=31$ system, but there is no clear way to identify the counting of levels and extract the associated edge theory. However, it can still be argued that there is a multi-branch structure, which is probably due to the $[\Phi_{3}]^{*}$ factor in the wave function. This is as expected due to the fact that this state exhibits multiple edge modes with opposite chiralities.

\section{Edge Physics of The Three Candidate States}

While these results make the BS state very plausible, its full confirmation will require further theoretical and experimental work. In particular, it would be crucial to ask if the BS description also captures the physics of the excitations. The excitations of this state are complicated, which can be seen as follows. In the standard CF theory, the excitations are created in the factor $\Phi_n$. Such an excitation in the BS wave function produces a quasiparticle of charge $1/8$. On the other hand an excitation in the Pf factor has a charge $3/16$. The most straightforward way to obtain these numbers is to note that a charge $3/8$ vortex at position $(U,V)$ on the sphere, obtained by multiplication with $\prod_i (u_i V-v_i U)$, can be combined with $\Phi_3^*$ and split into three quasiholes (each with charge $1/8$), or it can be combined with the Pf factor and split into two quasiholes (each with charge $3/16$). The object with smallest charge $1/16$ consists of one excitation of each type. We now ask, within an effective theory framework, what are the most relevant excitations for various kinds of tunnelings at the edge.

In the following, we perform an analysis of the Pf and the APf  3/8 states, and the results (correcting the analysis of Ref.~\cite{Mukherjee12}) are given in Table \ref{tab:exc}. The three states have very distinct electron structures. Thus, if one could tunnel an electron directly into the edge of the 3/8 FQHE state, one would be able to differentiate between these states from the characteristic exponent of the tunneling current. Unfortunately, an electron from the outside couples preferably to the edge of the {\em lowest} LL. The most relevant quasiparticles in all of these states have identical scaling dimensions. (We emphasize that these results are obtained under the assumption that the edge is in the universal regime where the two-terminal Hall conductance is universal and quantized. Outside this regime, the interaction between forward moving charged mode and backward moving neutral modes yields non-universal tunneling exponents, but with the price of non-universality of the two-terminal Hall conductance \cite{Hutasoit15}.) Therefore, it will not be possible to differentiate between these states using QPC tunneling experiment, unless somehow one is able to probe the less relevant quasiparticles. This and the recent progress in heat transport measurement \cite{Banerjee16} make the thermal Hall conductance measurement the most promising way to experimentally differentiate between these states. The thermal Hall conductance $\kappa_H=\partial J_Q/\partial T$, where $J_Q$ is the thermal energy current and $\partial T$ is the Hall temperature difference. It only depends on the central charge of the edge CFT and not on the interaction between the modes, which means that it takes a universal and quantized value. In units of $(\pi^2 k_B^2 /3h)T$, each chiral boson edge mode contributes one unit and each Majorana fermion mode 1/2 unit \cite{Kane97,Kitaev06}, with the sign depending on the direction of propagation. The BS state has two backward and one forward moving bosonic modes and one forward moving Majorana mode, producing $\kappa_H = -1/2$. On the other hand, the thermal Hall conductance for both Pfaffian and anti-Pfaffian states are positive, given by $\kappa_H = 5/2$ and $\kappa_H = 1/2$, respectively. We note that the thermal Hall conductance will also differentiate these non-Abelian states from the Abelian state of Fr{\"o}hlich {\em et al.} \cite{Frohlich97}, which has $\kappa_H=3$. (The lowest filled LL will contribute an additional $\kappa_H=2$ to all the states.) 

\begin{table}[h]
\begin{ruledtabular}
\begin{tabular}{ l | l  l  c}
  &  BS & Pf & APf \qquad \\
  \hline
  electron &  $\{\psi,1,\frac{13}{6}\}$ & $\{\mathds{1},1,\frac{3}{2}\}$ & $\{\mathds{1},1,\frac{5}{3}\}$ \\
  & & $\{\psi,1,\frac{5}{2}\}$ & $\{\psi,1,\frac{13}{6}\}$ \\
  \hline
  quasiparticle &  $\{\sigma,\frac{3}{16},\frac{7}{64}\}$ & $\{\sigma,\frac{3}{16},\frac{7}{64}\}$ & $\{\sigma,\frac{3}{16},\frac{7}{64}\}$\\
  & $\{\sigma,\frac{1}{8},\frac{29}{192}\}$ & $\{\mathds{1},\frac18,\frac{3}{16}\}$ & $\{\sigma,\frac{1}{16},\frac{29}{192}\}$ \\
  & & $\{\mathds{1},\frac38,\frac{3}{16}\}$ &
\end{tabular}
\end{ruledtabular}
\caption{The spectrum for three candidates of $\nu=2+3/8$ state, namely the Bonderson-Slingerland (BS), Pfaffian (Pf), and the anti-Pfaffian (APf) wave functions. Inside the curly brackets $\{\cdots \}$ the first entry denotes the sector of the Ising CFT the particle lives in, the second entry denotes the charge, and the last entry denotes the scaling dimension.} 
\label{tab:exc}
\end{table}

\subsection{The formalism}

Let us start with a quick summary of the $K$ matrix terminology, following Ref. \onlinecite{Moore98}. A quantum Hall edge with $n$ chiral bosonic modes is described by the Hamiltonian
\be
H_b = \frac{1}{4 \pi}\int dx \, V_{ij} \,\partial_x \phi_i \, \partial_x \phi_j, \label{eq:HamS1}
\ee
where $i = 1, \cdots, n$ and the bosons obey the following commutation relation
\be
[\phi_i(x),\phi_j(x')] = i \pi \left[ 
\left(K^{-1}\right)_{ij} {\rm sgn}(x-x') + \epsilon_{ij} \right],
\ee
with $K \in {\rm SL}(n, \mathbb{Z})$ a symmetric matrix and the anti-symmetric matrix
\bea
\epsilon_{ij}= 
\begin{cases}
    1,& \text{if }  i>j\\
    -1,              & \text{otherwise}
\end{cases}
\eea
ensures all electron vertex operators are mutually fermionic. The filling factor is given by
\be
\nu = t^T \cdot K^{-1} \cdot t,
\ee
where the entries of the charge vector $t$ are integers. The spectrum can then be constructed using vertex operators. A quasiparticle that is described by ${\cal O}_{\ell} = e^{i \ell_i \phi_i}$%, with integer $\ell_i$'s,
 has a charge $q_{\ell} = t^T \cdot K^{-1} \cdot \ell$ and its exchange statistics with respect to another quasiparticle  ${\cal O}_{\ell'} = e^{i \ell'_i \phi_i}$ (which can be itself) is given by $\theta_{\ell \ell'} = \pi \, \ell^T \cdot (K^{-1}+\epsilon) \cdot \ell'$. 

The edge theory can also have a neutral sector constructed from the chiral (or anti-chiral) part of a rational conformal field theory (CFT) and in that case, the quasiparticles will be described by products of primary operators of the rational CFT and the vertex operators of the bosonic sector. The charge of such a quasiparticle depends solely on the bosonic part but the exchange statistics will also have to include the exchange statistics of the operators of the rational CFT.

To construct the spectrum, one first finds all fermionic operators with charge equal to unity. These are the electron operators. Since any \emph{physical} correlation function containing an electron operator insertion must be a single-valued function of the position of the insertion, the rest of the spectrum can be constructed by finding all operators that are single-valued under the exchange with the electron operators \cite{Wen92b,Frohlich01}. In other words, we can find all physical quasiparticles by finding operators whose exchange statistics with respect to all electron operators are integer multiples of $\pi$.

In order to determine the scaling dimensions of the quasiparticles, we need to simultaneously diagonalize $K$ and $V$. First, let us consider a basis transformation $\phi' = M_1^{-1} \cdot \phi$, under which 
\be
K' = M_1^T \cdot K \cdot M_1 =    \begin{pmatrix} % or pmatrix or bmatrix or Bmatrix or ...
      \mathds{1}_{n_+} & 0 \\
      0 & -\mathds{1}_{n_-}  \\
   \end{pmatrix},
\ee
where $\mathds{1}_{n_{\pm}}$ is an $n_{\pm} \times n_\pm$ identity matrix and $n_+ + n_- = n$. Next, we can diagonalize $V'=M_1^T \cdot V \cdot M_1 $ by
\be
V'' = M_2^T \cdot M_1^T \cdot V \cdot M_1 \cdot M_2,
\ee
where $V''$ is a diagonal matrix and $M_2 \in SO(n_+,n_-)$ such that $K'' = K'$. (Note that $V$ is a symmetric matrix.) We can express the second basis transformation as $M_2 = B \cdot R$, where $R$ is an orthogonal matrix, \textit{i.e.}, the rotation, and $B$ is a positive matrix, \textit{i.e.}, the pure boost of Lorentz group. It turns out that the scaling dimension of an operator ${\cal O}_{\ell''}$ is given by 
\be
\Delta_{\ell''} = {\ell''}^T \cdot \Delta \cdot {\ell''}.
\ee
where
\be
\Delta = \frac{B^2}{2} . \label{eq:delta}
\ee 
Interestingly $\Delta$ also determines the two-terminal Hall conductance, which is given by
\be
\sigma_H = 2 \, {t''}^T \cdot \Delta  \cdot {t''}. \label{eq:Hallc}
\ee
Here, the two-terminal conductance is given by the current response to an electric field applied along the edge.

\subsection{Bonderson-Slingerland state}

Let us now look at the edge theory for Bonderson-Slingerland state. It is given by
\be
K =    \begin{pmatrix} % or pmatrix or bmatrix or Bmatrix or ...
      2 & 3 & 3 \\
      3 & 2 & 3 \\
     3 & 3 & 2
   \end{pmatrix}
\qquad {\rm and} \qquad t = \begin{pmatrix} % or pmatrix or bmatrix or Bmatrix or ...
      1 \\
      1 \\
      1
   \end{pmatrix},
\ee
along with the chiral (holomorphic) part of an Ising CFT. The latter has two primary operators $\psi$ and $\sigma$, whose exchange statistics are given by
\be
\theta_{\psi\psi}=\pi, \qquad \theta_{\psi \sigma}=\pi/2,
\ee
and scaling dimensions are given by
\be
\Delta_{\psi}=1/2, \qquad \Delta_{\sigma}=1/16.
\ee
The $K$ matrix has a close correspondence with the BS wave function: It is given by a $3\times 3$ matrix $K=-\mathds{1}+3C_3$, where the first term on the right hand side is the $K$ matrix of $\Phi_3^*$ and the second term, with $C_3$ being the pseudo-identity matrix of all ones, arises from the Jastrow factor $\prod_{j<k}(u_jv_k-u_kv_j)^3$. $V_{ij}$ denotes the strength of the short range interaction between the bosons, and $v$ is the velocity of the Majorana mode. 

We first note that all charge 1 vertex operators are mutually bosonic and thus, all electron operators must reside in the $\psi$-sector of the Ising CFT. It is instructive to parametrize the boost $B$ as 
\be 
B^2 =    \begin{pmatrix} % or pmatrix or bmatrix or Bmatrix or ...
      \gamma & \beta_1 \gamma & \beta_2 \gamma \\
      \beta_1 \gamma & 1+ \frac{\beta_1^2 \gamma^2}{\gamma+1} & \frac{\beta_1 \beta_2 \gamma^2}{\gamma+1} \\
      \beta_2 \gamma & \frac{\beta_1 \beta_2 \gamma^2}{\gamma+1} & 1+ \frac{\beta_2^2 \gamma^2}{\gamma+1}  
       \end{pmatrix},
\ee
where $\gamma = 1/\sqrt{1-\beta_1^2-\beta_2^2}$, in the basis where 
\be
K =  {\rm diag}(1,-1,-1)
\qquad {\rm and} \qquad t = (\sqrt{3/8},0,0)^T.
\ee
The two-terminal Hall conductance is then given by $\sigma_H = 3 \gamma/8$ and thus, we should look at the point $\gamma=1$, or $\beta_1=\beta_2=0$. This is the point in parameter space where the forward moving charged mode is decoupled from the two backward moving neutral modes.

A comment is in order. Since the regime where the Hall conductance is given by the ``correct" value is only a point in the parameter space, a mechanism where this point becomes an attractive renormalization group (RG) fixed point is desired. Such mechanism is given in Ref. \onlinecite{Kane94}. Regardless of the mechanism, away from this point both the two-terminal Hall conductance and the scaling dimensions of quasiparticles are not universal \cite{Hutasoit15}.

When $\gamma=1$, there are three electron operators that are most relevant. In the original basis,
they are given by 
\be
{\cal O}_{el}= \psi \, \exp[i \ell_i \phi_i],
\ee
where $\{\ell_i\}$ is the permutation of $\{2,3,3\}$, and their scaling dimension is 13/6. The most relevant quasiparticle in the $\mathds{1}$-sector of the Ising CFT is given by 
\be
{\cal O}_{\mathds{1}} = \exp\Big[i \left(\phi_1+\phi_2+\phi_3\right)\Big],
\ee
whose charge is 3/8 and scaling dimension is 3/16. Similarly, the most relevant quasiparticle in the $\psi$-sector is given by 
\be
{\cal O}_{\psi} = \psi \exp\Big[i \left(\phi_1+\phi_2+\phi_3\right)\Big].
\ee
Its charge is also 3/8 but the scaling dimension is 11/16. Lastly, the most relevant quasiparticle in the $\sigma$-sector is given by 
\be
{\cal O}_{\sigma} = \sigma \exp\left[i\frac{\phi_1+\phi_2+\phi_3}{2}\right],
\ee
whose charge is 3/16 and scaling dimension is 7/64. This quasiparticle will dominate the tunneling between the edges through a quantum point contact (QPC). 
The tunneling current is given by $I \sim V^{2g-1}$ and the tunnel conductance has a temperature dependence given by $\sigma_T \sim T^{2g-2}$, where $g=7/32$ is twice the scaling dimension.  

Since the scaling dimension of ${\cal O}_{\sigma}$ is smaller than that of the most relevant operators in the other sectors, let us also look at the second most relevant quasiparticles in the $\sigma$-sector. There are three of them and they are given by 
\be
{\cal O}_{\sigma,i} = \sigma \exp\left[i\frac{\phi_i}{2}\right],
\ee
where $i=1$, 2 or 3. Their charge is 1/8 and the scaling dimension is 29/192, which is still smaller than the scaling dimension of the most relevant operator in the $\mathds{1}$-sector. It is interesting to note that the quasiparticles with the smallest charge do not have the smallest scaling dimension. This is because the requirement that all operators that represent physical excitations must be single-valued under the exchange with the electron operators results in the quasiparticles with the smallest charge to depend not only on the forward moving charged mode but also on the neutral modes. 

The next quasiparticle also comes in triplet: 
\be
{\cal O}_{\sigma,i,j} = \sigma \exp\left[i\frac{\phi_i+\phi_j}{2}\right],
\ee
where $i\ne j$. Its charge is 1/8 and the scaling dimension is 1/6. 

\subsection{Pfaffian $3/8$}

The bosonic content of the edge theory of the 3/2 state in which the LL is full and the second LL has 1/2 state is described by the $K$ matrix $\begin{pmatrix} % or pmatrix or bmatrix or Bmatrix or ...
      1 & 0  \\
      0 & 2  \\
         \end{pmatrix}$. Composite-fermionization of this state adds 2 to each entry to produce the $K$ matrix for the 3/8 
Pfaffian state:
\be
K =    \begin{pmatrix} % or pmatrix or bmatrix or Bmatrix or ...
      3 & 2  \\
      2 & 4  \\
         \end{pmatrix}
\qquad {\rm and} \qquad t = \begin{pmatrix} % or pmatrix or bmatrix or Bmatrix or ...
      1 \\
      1 \\
   \end{pmatrix},
\ee
which is then accompanied with the chiral Ising CFT.  
There are charge-1 fermionic operators both in the $\mathds{1}$-sector and in the $\psi$-sector:
\begin{eqnarray}
 {\cal O}_{el,\mathds{1}} &=& \exp\Bigg[i \Big\{(3+2n)\phi_1+(2-4n) \phi_2\Big\} \Bigg] \\
 {\cal O}_{el,\psi} &=& \psi \exp\Bigg[i \Big\{(2+2n)\phi_1+(4-4n) \phi_2\Big\} \Bigg],
\end{eqnarray}
where $n \in \mathbb{Z}$ and the operators with the smallest scaling dimensions in each sector are given by $n=0$. The scaling dimensions of these most relevant operators are 3/2 for the $\mathds{1}$-sector and 5/2 for the $\psi$-sector. We note that even though all electrons in each sector are mutually fermionic, electrons from different sectors are mutually bosonic. This can be fixed by introducing an extra Klein factor.

Now, we can construct quasiparticle operators by finding all operators whose correlators are single-valued under an electron operator insertion. In other words, all operators whose exchange statitistics with all electron operators are integer multiples of $\pi$. In $\mathds{1}$-sector, there are two of the most relevant of such operators and they are given by 
\be
{\cal O}_{\mathds{1},\tfrac{1}{8}} = \exp [i \phi_2],~~{\cal O}_{\mathds{1},\tfrac{3}{8}} = \exp \Big[i \left(\phi_1+\phi_2\right)\Big]
\ee
They both have scaling dimension 3/16 but the former has charge 1/8 while the latter 3/8. Similarly, in $\psi$-sector, we have two most relevant operators
\be
{\cal O}_{\psi,\tfrac{1}{8}} = \psi \exp [i \phi_2],~{\cal O}_{\psi,\tfrac{3}{8}} =  \psi\exp \Big[i \left(\phi_1+\phi_2\right)\Big]
\ee
both with scaling dimension 11/16 but the former with charge 1/8 and the latter with charge 3/8.

In the $\sigma$-sector, the single-value criterion requires the coefficients in front of $\phi_i$'s to be odd multiples of half-integers. Therefore, the most relevant quasiparticle in that sector is unique and it is given by
\be
{\cal O}_{\sigma} = \sigma \exp \left(i \frac{\phi_1+\phi_2}{2}\right),
\ee
with charge 3/16 and scaling dimension 7/64. 

\subsection{Anti-Pfaffian $3/8$}

The bosonic content of the edge theory of the 3/2 APf state in which the lowest two LLs are full and the second LL has 1/2 state of holes is described by the $K$ matrix 
$\begin{pmatrix} % or pmatrix or bmatrix or Bmatrix or ...
      1 & 0 & 0  \\
      0 & 1 &0 \\
      0 & 0 & -2
         \end{pmatrix}$. Composite-fermionization of this state adds 2 to each entry to produce the $K$ matrix for the 3/8 
anti-Pfaffian state: 
\be
K=\begin{pmatrix} % or pmatrix or bmatrix or Bmatrix or ...
      3 & 2 & 2  \\
      2 & 3 &2 \\
      2 & 2 & 0
         \end{pmatrix}
\qquad {\rm and} \qquad t = \begin{pmatrix} % or pmatrix or bmatrix or Bmatrix or ...
      1 \\
      1 \\
      1
   \end{pmatrix}
\ee
It has two forward and one backward moving chiral boson modes. We make an $SL(3,\mathbb{Z})$ transformation with 
$\begin{pmatrix} -1& 1 & 1 \\ 3 & 0 & -2\\ 0 & 1 & 0 \end{pmatrix}$ to bring it into the
equivalent form  (which we prefer because it has integer eigenvalues) 
\be
K =    \begin{pmatrix} % or pmatrix or bmatrix or Bmatrix or ...
      2 & 3 & 3 \\
      3 & 3 & 2 \\
      3 & 2 & 3
   \end{pmatrix}
\qquad {\rm and} \qquad t = \begin{pmatrix} % or pmatrix or bmatrix or Bmatrix or ...
      1 \\
      1 \\
      1
   \end{pmatrix}.
\ee
This, along with the anti-chiral part of an Ising CFT:
\be
\theta_{\psi\psi}=-\pi, \qquad \theta_{\psi \sigma}=-\pi/2,
\ee
then describe the anti-Pfaffian state. We note that there is an $SU(2)$ symmetry, with $\phi_{2,3}$ transforming as a doublet. Here, we have a counter-propagating bosonic mode and thus, in most part of the parameter space, observables such as the two-terminal Hall conductance and scaling exponents are not universal. Let us parametrize the boost $B$ as 
\be 
B^2 =    \begin{pmatrix} % or pmatrix or bmatrix or Bmatrix or ...
       1+ \frac{\beta_1^2 \gamma^2}{\gamma+1}  & \frac{\beta_1 \beta_2 \gamma^2}{\gamma+1} & \beta_1 \gamma \\
    \frac{\beta_1 \beta_2 \gamma^2}{\gamma+1}   & 1+ \frac{\beta_2^2 \gamma^2}{\gamma+1} &\beta_2 \gamma  \\qo
     \beta_1 \gamma  & \beta_2 \gamma  &  \gamma
       \end{pmatrix},
\ee
where $\gamma = 1/\sqrt{1-\beta_1^2-\beta_2^2}$, in the basis where 
\be
K =  {\rm diag}(1,1,-1)
\qquad {\rm and} \qquad t = (\sqrt{3/8},0,0)^T.
\ee
The two-terminal Hall conductance is then given by 
\be
\sigma_H = \frac{3}{8} \left(1+ \frac{\beta_1^2 \gamma^2}{\gamma+1}\right),
\ee
and thus, we should look at the point $\beta_1=0$. This is the point in parameter space where the forward moving charged mode is decoupled from the backward moving neutral mode. Since $\beta_1=0$, $|\beta_2|<1$ is the universal regime, for simplicity, we can set $\beta_2=0$ \cite{Hutasoit15}.

As is in the case of Pfaffian 3/8, here we also have charge-1 fermionic operators in both the $\mathds{1}$-sector and $\psi$-sector. In the $\mathds{1}$-sector, they take the form:
\begin{eqnarray*}
{\cal O}_{el,\mathds{1}}&=&\exp\Bigg[i \Big\{(7-2m-2n)\phi_1+(2m+1) \phi_2+ 2n \phi_3\Big\} \Bigg]  \\
{\cal O}_{el,\mathds{1}}&=&\exp\Bigg[i \Big\{(7-2m-2n)\phi_1+2m \phi_2+(2n+1)\phi_3\Big\} \Bigg]
\end{eqnarray*}
while in $\psi$-sector, they take the following form:
\bea
{\cal O}_{el,\psi} &=& \psi \exp\Bigg[i \Big\{(6-2m-2n)\phi_1+(2m+1) \phi_2 \nonumber \\
& & \qquad \qquad +\, (2n+1) \phi_3\Big\} \Bigg]  \nonumber \\
{\cal O}_{el,\psi} &=& \psi \exp\Bigg[i \Big\{(8-2m-2n)\phi_1+2m \phi_2+2n\phi_3\Big\} \Bigg]. \nonumber \\
\eea
The most relevant electrons are then a doublet in the $\mathds{1}$-sector
\begin{eqnarray}
{\cal O}_{el,\mathds{1}} &=& \exp\Big[i \left(3\phi_1+3 \phi_2+ 2 \phi_3\right) \Big]  \\
{\cal O}_{el,\mathds{1}} &=& \exp\Big[i \left(3\phi_1+2\phi_2+3\phi_3\right) \Big]
\end{eqnarray}
with scaling dimension 5/3, while in the $\psi$-sector
\be
{\cal O}_{el,\psi} = \psi \exp\Big[i \left(2\phi_1+3 \phi_2+ 3 \phi_3\right) \Big], 
\ee
with scaling dimension 13/6.

Now, let us look at the quasiparticle content. In the $\mathds{1}$-sector, the most relevant quasiparticle is
\be
{\cal O}_{\mathds{1}} = \exp\Big[i \left(\phi_1+\phi_2+ \phi_3\right) \Big],
\ee
with charge 3/8 and scaling dimension 3/16. Similarly, in the $\psi$-sector, we have
\be
{\cal O}_{\psi} = \psi\exp\Big[i \left(\phi_1+\phi_2+ \phi_3\right) \Big],
\ee
with the same charge but scaling dimension 11/16.

For the $\sigma$-sector, the most relevant quasiparticle is given by
\be
{\cal O}_{\sigma} = \sigma \exp\left[i \frac{\phi_1+\phi_2+ \phi_3}2 \right],
\ee
with charge 3/16 and scaling dimension 7/64. The next most relevant quasiparticle in this sector is
\be
{\cal O}_{\sigma} = \sigma \exp\left[i \frac{\phi_1}2 \right],
\ee
with charge 1/16 and scaling dimension 29/192, which is smaller than the scaling dimension of the most relevant quasiparticle in the $\mathds{1}$-sector.

\section{Conclusion}

In summary, by studying several candidate states we have shown that the state at $\nu=2+3/8$ is probably a realization of a new class of Bonderson-Slingerland FQH states. The energy spectrum on torus provides some evidence for 24-fold ground state degeneracy, and the entanglement spectrum suggests that the edge contains multiple branches, but study of larger systems, currently not possible, would be necessary for more detailed and definitive information. We have further shown, using an effective edge theory, that the BS structure of this state can be confirmed either by tunneling of an electron into the 3/8 edge or by a measurement of the thermal Hall conductance. 

\begin{acknowledgments} 

The authors are grateful to Csaba T\H oke for useful discussions. We also thank Parsa Bonderson and his collaborators for comments on the manuscript and for sharing their earlier unpublished calculation of the overlap of the BS state with the exact Coulomb state that is consistent with ours. This work was supported by the U. S. National Science Foundation Grant no. DMR-1401636 (A.C.B., J.K.J.), Polish NCN Grant No. 2014/14/A/ST3/00654 (A.W.), European Union Project SIQS (Y.H.W.), the Foundation for Fundamental Research on Matter (FOM), the Netherlands Organization for Scientific Research (NWO/OCW), and an ERC Synergy Grant (J.H.,V.C.). We thank Research Computing and CyberInfrastructure at Pennsylvania State University (supported in part through instrumentation funded by the National Science Foundation through Grant No. OCI-0821527), and Wroc\l{}aw Centre for Networking and Supercomputing and Academic Computer Centre CYFRONET, both parts of PL-Grid Infrastructure.  

\end{acknowledgments}

%\bibliography{../../../Latex-Revtex-etc./biblio_fqhe}
\bibliography{biblio_fqhe}
\bibliographystyle{apsrev}
\end{document}